\documentclass[11pt]{article}
\usepackage{amssymb,latexsym,amsmath} 
\usepackage{amsfonts, graphics}
\newcommand{\R}{\mathbb R}

\def\open#1{\setbox0=\hbox{$#1$}
\baselineskip = 0pt
\vbox{\hbox{\hspace*{0.4 \wd0}\tiny $\circ$}\hbox{$#1$}}
\baselineskip = 11pt\!}
\def\fn{\open{f}}

\def\dt{\partial_t}
\def\dx{\partial_x}
\def\dv{ \partial_v }

\def\be{\begin{equation}}
\def\ee{\end{equation}}
\def\bea{\begin{eqnarray}}
\def\eea{\end{eqnarray}}
\def\beas{\begin{eqnarray*}}
\def\eeas{\end{eqnarray*}}

\def\supp{\mathrm{supp}\,}

\begin{document}

\newtheorem{theorem}{Theorem}[section]
\renewcommand{\thetheorem}{\arabic{section}.\arabic{theorem}}
\newtheorem{definition}[theorem]{Definition}
\newtheorem{proposition}[theorem]{Proposition}
\newtheorem{example}[theorem]{Example}
\newtheorem{remark}[theorem]{Remark}
\newtheorem{cor}[theorem]{Corollary}
\newtheorem{lemma}[theorem]{Lemma}

\title{Gravitational collapse and the formation of black holes 
       for the spherically symmetric Einstein-Vlasov system}

\author{H{\aa}kan Andr\'{e}asson\thanks{Support by the 
        Institut Mittag-Leffler (Djursholm, Sweden) is gratefully 
        acknowledged.}\\
        Mathematical Sciences\\
        Chalmers University of Technology\\
        G\"{o}teborg University\\
        S-41296 G\"oteborg, Sweden\\
        email: hand@math.chalmers.se\\
        \ \\
        Markus Kunze\\
        Fachbereich Mathematik\\
        Universit\"at Duisburg-Essen\\
        D-45117 Essen, Germany\\
        email: markus.kunze@uni-due.de\\
        \ \\
        Gerhard Rein\\
        Mathematisches Institut der Universit\"at Bayreuth\\
        D-95440 Bayreuth, Germany\\
        email: gerhard.rein@uni-bayreuth.de}
\date{{\em Dedicated to Prof.~W.~A.~Strauss on the occasion of his 70th birthday}}
\maketitle

\begin{abstract}
We review results on the spherically symmetric, asymptotically flat
Einstein-Vlasov system. We focus on a recent result
where we found explicit conditions on the initial data
which guarantee the formation of a black hole in the evolution. 
Among these data there are data such that the corresponding solutions 
exist globally in Schwarzschild coordinates. 
We put these results into a more general context, and we include 
arguments which show that the spacetimes we obtain satisfy
the weak cosmic censorship conjecture and 
contain a black hole in the sense of suitable
mathematical definitions of these concepts which are available in 
the literature.
\end{abstract}
\tableofcontents
\section{Some general remarks on gravitational collapse}
\setcounter{equation}{0}
We start these notes with some general, informal, and in part historical remarks
on concepts related to the phenomenon of relativistic gravitational collapse.
 
Shortly after A.~Einstein published his theory of 
general relativity \cite{Einst1,Einst2}, K.~Schwarzschild showed that the following
metric solves the corresponding field equations in vacuum \cite{Schw}:
\be \label{scm}
ds^2 = -\left(1-\frac{2 M}{r}\right)\, dt^2 + \left(1-\frac{2 M}{r}\right)^{-1}dr^2\,
+ r^2(d\theta^2 + \sin^2 \theta d\varphi^2).
\ee
Here $M>0$ is a parameter, $t\in \R$ is a time coordinate, the
spacetime is spherically symmetric, and the polar angles $\theta$ and $\varphi$
coordinatize the surfaces of constant $t$ and $r>0$.
The latter are the orbits
of $\mathrm {SO}(3)$ which acts isometrically on this spacetime, and
$4 \pi r^2$ is the area of these surfaces. 
The part of this metric with $r>2 M$ can be thought of as representing 
the gravitational field outside a static, spherically symmetric body 
of mass $M$ and radius larger than $2 M$. With this interpretation 
in mind the radii $r=2 M$ and $r=0$, where the metric looks singular, 
lie within the matter where the metric does not apply anyway so that 
one need not worry about this singular behavior. 
However, using the new time coordinate
\be \label{eddfink}
\tilde t := t + r^\ast\ \mbox{where}\ r^\ast := r + 2 M \ln (r-2 M),
\ee
the metric takes the form
\be \label{sceddfink}
ds^2 = -\left(1-\frac{2 M}{r}\right)\, {d\tilde t}^2 + 2 d\tilde t\,dr\,
+ r^2(d\theta^2 + \sin^2 \theta d\varphi^2),
\ee 
which extends smoothly through $r=2 M$, i.e., $r=2 M$ is only
a {\em coordinate singularity} in (\ref{scm}). On the other hand,
the so-called Kretschmann scalar
\be \label{kretschmann}
K := R^{\alpha \beta \gamma \delta} R_{\alpha \beta \gamma \delta}
\ee
derived from the Riemann curvature tensor 
$R_{\alpha \beta \gamma}^{\phantom{\alpha \beta \gamma}\delta}$
blows up like $M^2/r^6$ as $r\to 0$; Greek indices are running from 
$0$ to $3$ and are summed over if they appear as both lower and upper
indices in the same expression. Since $K$ is a scalar quantity, 
its values do not change
under a change of coordinates. The singularity at $r=0$ is a genuine
feature of the Schwarzschild spacetime which cannot be cured away by
a more judicious choice of coordinates, it is a {\em spacetime singularity}
where the structure of spacetime itself breaks down.
Although the surface $r=2 M$ is not singular there is something special
about it, as can be seen from (\ref{sceddfink}). The line
$r=2 M,\ \theta =\pi/2,\ \varphi=0$ represents the world line of a massless
particle (a photon) moving radially outward, away from the origin. Since 
no material particle can move faster than light, particles and photons can 
only pass through the surface $r= 2 M$ inward, but can never leave the region
$r< 2 M$ once they are inside. Such a surface was later termed an 
{\em event horizon}. One should keep in mind from the above that by switching 
to different coordinates one was able to extend the spacetime beyond
the region covered by Schwarzschild coordinates, even though
in the latter coordinates the metric blows up at the boundary $r=2 M$.

While the part with $r > 2 M$ of the Schwarzschild metric was successfully 
used to explain for example
the perihelion precession of Mercury in the solar system, it was argued that
no conceivable physical process could compress an amount of matter so much
that its mass $M$ would all be contained inside the region $r< 2 M$, 
and so the irritating behavior at the surface $r=2 M$ and also the break down
of the geometry of spacetime at $r=0$ was discarded as unphysical.

But in 1939 J.~R.~Oppenheimer and H.~Snyder \cite{OS} constructed a semi-explicit,
time dependent solution of the Einstein field equations
where a homogeneous spherically symmetric ball of dust, i.e., of a 
fluid with pressure zero, collapses until all the mass is within the region
$r < 2 M$, and it continues to collapse until the scalar
curvature of spacetime blows up at the center of symmetry.
Although this proved that an event horizon can evolve out of completely 
regular initial data, several decades passed before such structures
became accepted as potentially relevant from the physics point of view
and J.~A.~Wheeler coined the name {\em black hole} for them.

Today, there are many astronomical observations for which the currently
best explanation involves a black hole. For example, black holes
of the order of $10^6$--$10^9$ solar masses are believed to reside in the centers
of many galaxies, including the Milky Way. In spite of the increasing relevance
of black holes as real astrophysical objects many important basic questions 
about gravitational collapse are still open. The most prominent of these is 
the {\em cosmic censorship conjecture}. 

To see the issue here, we notice that
in the Oppenheimer-Snyder example the spacetime singularity
which forms at $r=0$ is hidden behind the event horizon so that
it cannot be seen or in any other way be experienced by observers outside 
the event horizon. The same is true for the spacetime singularity in the
Schwarzschild spacetime. 
In the 1960s R.~Penrose \cite{pen} proved that a spacetime 
singularity forms in the gravitational collapse of a not necessarily symmetric
star made up of ``reasonable'' matter, i.e., matter which satisfies the
strong energy condition, provided a {\em closed trapped surface} forms.
This is a closed, two-dimensional, spacelike surface with the property that
the null geodesics, i.e., the light rays, which start perpendicular to
the surface decrease its surface area both when followed inward and outward
from the surface; for a precise definition cf.\ \cite[9.5]{Wald}. 
The surfaces of constant $\tilde t$ and $r$ with $r<2 M$ in (\ref{sceddfink})
are trapped, and we note that Schwarzschild coordinates cannot 
cover regions of spacetime which contain trapped surfaces.
Since trapped surfaces are stable under small perturbations
of the spacetime, Penrose's result showed that the formation of spacetime singularities
is not restricted to spherically 
symmetric, especially constructed or isolated examples but is a genuine,
stable feature of spacetimes. However, his result gives little information
about the geometric structure of a spacetime with such a singularity,
in particular it does not provide an event horizon surrounding the
singularity.
The existence of a {\em naked singularity}
which by definition is not hidden behind an event horizon 
would violate predictability as it would not be possible 
to predict from initial data what an observer would
see if he could observe a singularity. Hence
Penrose formulated the cosmic censorship conjecture which demands that any
singularity 
arising in the gravitational collapse of {\em generic} regular initial data
is hidden behind an event horizon; exceptional data leading to naked
singularities are required to form a ``null set'' in some suitable sense.
The above is an informal statement of the so-called {\em weak cosmic censorship
conjecture} \cite[12.1]{Wald}. It would in particular guarantee that 
predictability holds at least in the region outside the event horizon.
In the strong version no observer is allowed to observe a singularity.
For a mathematical definition and discussion of the weak and strong
cosmic censorship conjectures we refer to \cite{Chr99}.

An important example where naked 
singularities do form for a null set of data while cosmic
censorship holds for generic data,
is the spherically symmetric Einstein-scalar field system 
which was investigated by D.~Christodoulou, cf.\ \cite{Chr94,Chr99a}.
A massless scalar field or dust as employed by Oppenheimer and
Snyder and later also by Christodoulou \cite{Chr84} are but two
possibilities for modeling matter in gravitational collapse.
In the present notes we discuss results where a collisionless gas as 
described by the Vlasov equation is used as matter model, a model which
we consider particularly suitable for this purpose from a mathematics point 
of view and which is well motivated from an astrophysics point of view,
cf.\ \cite{BT}.

\section{The Einstein-Vlasov system}
\setcounter{equation}{0}

Consider a smooth spacetime manifold $M$
equipped with a  Lorentzian metric $g_{\alpha \beta}$
with signature $(-{}+{}+{}+)$.
The Einstein equations read
\begin{equation} \label{feqgen}
G_{\alpha \beta} = 8 \pi T_{\alpha \beta},
\end{equation}
where $G_{\alpha \beta}$ is the Einstein tensor, a non-linear second order
differential expression in the metric $g_{\alpha \beta}$, and
$T_{\alpha \beta}$ is the energy-momentum tensor given by the
matter content (or other fields) of the spacetime. To obtain a closed system,
the field equations (\ref{feqgen}) have to be supplemented by
evolution equation(s) for the matter and
the definition of $T_{\alpha \beta}$ in terms of the matter and the metric.

We choose as matter model a collisionless gas. In order to
write down an evolution equation for the number density of the particles on
phase space we recall that the world line of a single test particle
on $M$ obeys the geodesic equation
\begin{equation} \label{chargen}
\dot x^\alpha = p^\alpha,\ 
\dot p^\alpha = - \Gamma^\alpha_{\beta \gamma} p^\beta p^\gamma,
\end{equation}
where $x^\alpha$ denotes general coordinates on $M$, 
$p^\alpha$ are the corresponding canonical momenta,
$\Gamma^\alpha_{\beta \gamma}$ are the Christoffel symbols
induced by the metric $g_{\alpha \beta}$,
and the dot indicates differentiation with respect to an affine
parameter, i.e., with respect to proper time along the
world line of the particle. 
We assume that all the particles have the same
rest mass, normalized to unity, and move forward in time.
Hence, their number density $f$ is a non-negative function
supported on the mass shell
\[
PM := \left\{ g_{\alpha \beta} p^\alpha p^\beta = -1,\ p^0 >0 \right\},
\]
a submanifold of the tangent bundle $TM$ of the spacetime manifold $M$
which is invariant under the geodesic flow.
Letting Latin indices range from $1$ to $3$
we use coordinates $(t,x^a)$ with zero shift which implies that
$g_{0a}=0$. On the mass shell $PM$ the variable $p^0$ then becomes 
a function of the remaining variables $(t,x^a,p^b)$:
\be \label{p0def}
p^0 = \sqrt{-g^{00}} \sqrt{1+g_{ab}p^a p^b}.
\ee
Since the particles move like test particles in the given metric,
their number density
$f=f(t,x^a,p^b)$ is constant along the geodesics, and it satisfies
the Vlasov equation
\begin{equation} \label{vlgen}
\partial_t f + \frac{p^a}{p^0}\,\partial_{x^a} f
-\frac{1}{p^0}\,\Gamma^a_{\beta \gamma} p^\beta p^\gamma\,\partial_{p^a} f = 0.
\end{equation}
The energy-momentum tensor is given by
\begin{equation} \label{emtvlgen}
T_{\alpha \beta}
=\int p_\alpha p_\beta f \,|g|^{1/2} \,\frac{dp^1 dp^2 dp^3}{-p_0},
\end{equation}
where $|g|$ denotes the modulus of the determinant of the metric,
and indices are raised and lowered using the metric, i.e.,
$p_\alpha = g_{\alpha \beta}p^\beta$.
The system (\ref{feqgen}), (\ref{vlgen}), (\ref{emtvlgen})
is the Einstein-Vlasov system in general coordinates.
An introduction to relativistic kinetic theory and the Einstein-Vlasov 
system can be found in \cite{And05,Rend97}.
The Vlasov equation is widely used as a matter model
in astrophysics, to describe galaxies or globular clusters \cite{BT}.
Such systems are usually dealt with as isolated systems in an otherwise
empty universe which in our context means that the spacetime is asymptotically
flat.

Let us for a moment consider a distribution function of the form
\be \label{dustf}
f(t,x^a,p^b) = - u_0\, |g|^{-1/2} \rho(t,x^a)\, \delta(p^b-u^b(t,x^a)),
\ee
where $\rho = \rho(t,x^a)$ is a scalar function on spacetime,
$\delta$ is the Dirac $\delta$-distribution, and
$u^\beta = u^\beta(t,x^a)$ takes values in the mass shell so that
$u_0$ is determined by $u^b$, cf.\ (\ref{p0def}). 
Then the macroscopic quantities
$\rho$ and $u^a$ together with the metric satisfy
the Einstein-Euler system for a perfect fluid with
pressure zero, a matter model referred to as dust
in the first section and used by Oppenheimer and Snyder;
$\rho$ is the mass-energy density of the dust-fluid and $u^\beta$
its four-velocity. It should be stressed that although formally
the Einstein-dust system can be viewed as a special case of
the Einstein-Vlasov system we use the term Vlasov matter
exclusively for genuine (and usually smooth) distribution
functions on the mass shell $PM$. 

Before we proceed a few advantageous features of the Einstein-Vlasov
system are worth to be pointed out. Firstly, if the metric
and therefore the Christoffel symbols are given, the evolution
equation for the matter, i.e., the Vlasov equation, does not
produce any singularities by itself; it is---for a given
metric---indeed a linear first order conservation law which can
be solved by the method of characteristics. This situation is 
different if the matter is described as a fluid, and singularities
induced by the matter model can prevent one from analyzing the formation
of event horizons and true spacetime singularities. Secondly,
in the Newtonian limit the Einstein-Vlasov system turns into
the Vlasov-Poisson system \cite{RR2,Rend94}
for which there is a global existence
result for general, regular data, cf.\ \cite{LP,Pf,Rein07}.
Hence any singularity in the solutions of the Einstein-Vlasov
system should have its origin in some relativistic effect;
for a fluid this is again different. Beside these
mathematical properties the Vlasov equation also has a clear
physical interpretation and motivation, as pointed out above.

The questions raised in the previous section are at present out of reach 
of a rigorous mathematical treatment, unless simplifying symmetry
assumptions are made. Hence we will consider the Einstein-Vlasov system
under the assumption of spherical symmetry.
Notice that the investigations which we cited above and where
dust or a massless scalar field are used as matter models
employ the same symmetry assumption.
We use Schwarzschild coordinates $(t,r,\theta,\varphi)$ and write
the metric in the form
\[
ds^2=-e^{2\mu(t,r)}dt^2+e^{2\lambda(t,r)}dr^2+
r^2(d\theta^2+\sin^2\theta\,d\varphi^2);
\]
as to the range and meaning of these coordinates we refer
to the previous section, cf.\ (\ref{scm}).
Asymptotic flatness means that the metric quantities $\lambda$ and $\mu$
have to satisfy the boundary conditions
\begin{equation}\label{boundcinf}
\lim_{r\to\infty}\lambda(t, r)=\lim_{r\to\infty}\mu(t, r)=0
\end{equation}
so that as $r\to \infty$ the metric approaches the flat
Minkowski metric, written in polar coordinates. In addition
we impose the boundary condition
\begin{equation}\label{boundc0}
\lambda(t, 0)=0
\end{equation}
in order to guarantee a regular center.
Since polar coordinates sometimes induce artificial coordinate
singularities at $r=0$, it is convenient 
to introduce the corresponding Cartesian coordinates
\[
x = (x^1,x^2,x^3) = 
r (\sin \theta \cos \varphi,\sin \theta \sin \varphi,
\cos \theta).
\]
If $p=(p^1,p^2,p^3)$ denotes the corresponding canonical momenta,
then
\[
p_0 = - e^\mu \sqrt{1+|p|^2 (e^{2\lambda} -1) 
+ \left(\frac{x\cdot p}{r}\right)^2},
\]
where $|p|^2 = (p^1)^2 + (p^2)^2 +(p^3)^2$ and 
$x\cdot p = x^1 p^1 + x^2 p^2 + x^3 p^3$. Since this 
quantity appears in the formula for the components of
the energy-momentum tensor which in turn appear as
source terms in the field equations, it is preferable
to use the non-canonical momentum variables
\[
v^a = p^a + (e^\lambda -1)\frac{x\cdot p}{r} \, \frac{x^a}{r},\ a=1,2,3.
\]
In these variables,
\[
p_0 = - e^\mu \sqrt{1+|v|^2},
\]
and $f$ is spherically symmetric iff
\[
f(t,x,v) = f(t,Ax,Av),\ x, v \in \R^3,\ A \in \mathrm{SO}\,(3).
\]
The spherically symmetric, asymptotically flat Einstein-Vlasov system
takes the following form: 
\be \label{vlasov}
\dt f + e^{\mu - \lambda}\frac{v}{\sqrt{1+|v|^2}}\cdot \dx f -
\left( \partial_t\lambda \frac{x\cdot v}{r} + e^{\mu - \lambda} \partial_r\mu
\sqrt{1+|v|^2} \right) \frac{x}{r} \cdot \dv f =0,
\ee
\be 
e^{-2\lambda} (2 r \partial_r\lambda -1) +1 
=
8\pi r^2 \rho , \label{ein1}
\ee
\be
e^{-2\lambda} (2 r \partial_r\mu +1) -1 
= 
8\pi r^2 p, \label{ein2} 
\ee
\be
\partial_t\lambda = 
- 4 \pi r e^{\lambda + \mu} j, \label{ein3}
\ee
\be
e^{- 2 \lambda} \left(\partial_r^2\mu + (\partial_r\mu - \partial_r\lambda)
(\partial_r\mu + \frac{1}{r})\right)
- e^{-2\mu}\left(\partial_t^2\lambda + 
\partial_t\lambda \, (\partial_t\lambda - \partial_t\mu)\right) 
=
4 \pi q, \label{ein4}
\ee
where
\bea
\rho(t,r) 
&=& 
\rho(t,x) = \int \sqrt{1+|v|^2} f(t,x,v)\,dv ,\label{r}\\
p(t,r) 
&=& 
p(t,x) = \int \left(\frac{x\cdot v}{r}\right)^2
 f(t,x,v)\frac{dv}{\sqrt{1+|v|^2}}, \label{p}\\
j(t,r) 
&=& 
j(t,x) = \int \frac{x\cdot v}{r} f(t,x,v) dv, \label{j}\\
q(t,r) 
&=& 
q(t,x) = \int \left|{\frac{x\times v}{r}}\right|^2
f(t,x,v) \frac{dv}{\sqrt{1+|v|^2}}. \label{q}
\eea
For a detailed derivation of these
equations we refer to \cite{Rein95}.
It should be noted that in this formulation no raising and
lowering of indices using the metric appears anywhere.
It is a completely explicit system of PDEs where $x,v \in \R^3$,
$x\cdot v$ denotes the Euclidean scalar product, and 
$|v|^2 = v\cdot v$.

For the spherically symmetric Einstein-dust equations
Christodoulou \cite{Chr84} showed that cosmic censorship
is violated. Indeed, not even a suitable smallness condition
on the initial data prevents the formation of naked singularities
for dust. As we shall see in the next section this is different
for the Vlasov matter model, provided we have a genuine,
smooth distribution function with respect to $x$ and $v$.

The goal for the spherically symmetric, asymptotically flat 
Einstein-Vlasov system is to show that for all (or at least
for all generic) regular initial data
the corresponding solution is either global in the sense that the spacetime
is singularity-free or the solution undergoes a gravitational collapse
in which a spacetime singularity forms which is hidden behind
an event horizon. Of course the ultimate goal would be to prove this
for general, not necessarily symmetric data, but let's be modest for now. 
So far all analytical and numerical results support the
conjecture that the above is indeed true and that in particular
the spherically symmetric, asymptotically flat
Einstein-Vlasov system satisfies the weak cosmic censorship
conjecture. An existence result for singularity-free solutions
for restricted, small data has been known for some time
and is reviewed in the next section. A class of data which lead
to gravitational collapse as described above has been established more
recently, and this result is discussed in Section~\ref{seccollapse}.

\section{Local and global existence results} \label{seclocglob}
\setcounter{equation}{0}

In this section we review a number of results
from the literature for the spherically symmetric, asymptotically 
flat Einstein-Vlasov system. They serve as either background
or counterparts to the results on gravitational collapse which
we state and discuss in the next section.

\subsection{Local existence and continuation}
 
Due to our choice of $v$ as a non-canonical momentum variable
the system (\ref{vlasov})--(\ref{q}) has the following nice feature.
If a distribution function $f$ is given, then the source terms
$\rho$, $p$, $j$, and $q$ can be computed from it without
reference to the metric. Given $\rho$ and observing the boundary
condition (\ref{boundc0}) the field equation (\ref{ein1})
can be integrated to yield
\be \label{e2lamb}
e^{-2\lambda}=1-\frac{2m}{r},
\ee
where the quasi-local mass $m$ is given by
\be \label{qlm}
m(t,r) := 4\pi \int_0^r \rho(t,\eta)\,\eta^2 d\eta.
\ee
Given $p$ and $\lambda$ the field equation 
(\ref{ein2}) together with the boundary condition
(\ref{boundcinf}) determines $\mu$:
\be \label{mu}
\mu(t,r) = \exp\left(-\int_r^\infty e^{2\lambda(t,\eta)}\, 
\left(\frac{m(t,\eta)}{\eta^2} + 4 \pi \eta p(t,\eta)\right)\,d\eta \right).
\ee
Finally, if both
$\lambda$ and $\mu$ are given and sufficiently regular,
then $f$ is determined from its initial data by the
method of characteristics:
\be \label{fviachar}
f(t,x,v) = \fn\, ((X,V)(0,t,x,v))
\ee
where $(X,V)(s,t,x,v)$ is the solution of the characteristic 
system
\be \label{charsys}
\dot x = e^{\mu - \lambda}\frac{v}{\sqrt{1+|v|^2}},\
\dot v = -
\left( \partial_t\lambda \frac{x\cdot v}{r} + e^{\mu - \lambda} \partial_r\mu
\sqrt{1+|v|^2} \right) \frac{x}{r}
\ee
of the Vlasov equation (\ref{vlasov}) satisfying
$(X,V)(t,t,x,v) = (x,v)$, and $\fn\,$ is the prescribed
data at time $t=0$. Notice that the characteristics are
now parameterized by coordinate time instead of proper time
as in (\ref{chargen}).

The iterative scheme indicated above can be used to prove
a local existence and uniqueness theorem together with an
extension criterion.
\begin{theorem} \label{le}
Let $\fn \in C^1_c(\R^6)$ be non-negative, spherically symmetric, 
and such that for the induced quasi-local mass,
\begin{equation} \label{init2mor}
2 \open{m}\,(r) / r < 1,\ 
r > 0.
\end{equation}
Then there exists a unique regular solution
$f$ of the asymptotically flat, spherically symmetric Einstein-Vlasov system
with $f(0)= \fn$ on a maximal interval of
existence $[0,T[$ with $T>0$.
If
\[
\sup \Bigl\{ |v| \mid (t,x,v) \in \supp f,\ 0\leq t < T \Bigr\} < \infty
\]
then $T = \infty$.
\end{theorem}
This result was first established in \cite{RR1}, see also \cite{Rein95}.
Some comments are in order. 

Functions in $C^1_c(\R^6)$ are by definition continuously differentiable and
compactly supported. A solution is called {\em regular} if the derivatives
which appear in the system exist in the classical sense and are continuous;
for the precise definition we refer to \cite{Rein95}.

The restriction (\ref{init2mor}) on the initial data
is necessary in view of (\ref{e2lamb}) because that equation defines
$\lambda$ only as long as $2 m (t,r)/r < 1$ so we have to require this for
the initial data. This is related to the fact that, as noted in the first section, 
Schwarzschild coordinates cannot cover regions of spacetime which
contain trapped surfaces.

A local existence and uniqueness result for the
Einstein-Vlasov system without a symmetry assumption was established
by Y.~Choquet-Bruhat \cite{CB}. However, in order to extend a local
solution to a global one based on the latter result one would have to
control high order Sobolev norms of the solution. The extension criterion
provided in Theorem~\ref{le} is much less demanding and forms
a more convenient starting point for investigating global properties
of solutions.
 
The major simplification of the system due to the symmetry assumption
is the fact that for given source terms the metric is completely determined 
by the two constraint equations (\ref{ein1}) and (\ref{ein2})
which are ordinary differential equations in $r$. 
The metric has no independent degrees of freedom, and this rules out
gravitational radiation, much like the Vlasov-Maxwell system
being reduced to the Vlasov-Poisson system if spherical symmetry
is assumed. 
But if $\lambda$ is defined by (\ref{e2lamb}), 
then it becomes rather unpleasant to control
$\partial_t\lambda$ which appears in the Vlasov equation and to make sure
that the condition $2 m (t,r)/r < 1$ is preserved. In \cite{Rein95}
this was resolved by considering first a modified system where
(\ref{ein4}) was left out and $\partial_t\lambda$ was replaced by a quantity
$\tilde \lambda$ defined by (\ref{ein3}). Then one can show a posteriori
that indeed $\partial_t\lambda = \tilde \lambda$ and that (\ref{ein4})
holds as well. Eqn.\ (\ref{ein3}) can then be used to control $\partial_t\lambda$
and therefore $\lambda$ in terms of 
\[
P(t) := \sup \Bigl\{ |v| \mid (s,x,v) \in \supp f,\ 0\leq s\leq t \Bigr\};
\]
notice that by (\ref{fviachar}) the function $f(t)$ is bounded so that
the source terms $\rho,p,j,q$ are bounded by powers of $P$.

Eqn.\ (\ref{ein4}) also has its role to play. To establish the
convergence of the iterative scheme one needs uniform bounds
on certain second order derivatives of the metric coefficients.
But since in (\ref{ein1})--(\ref{ein3}) the first order derivatives
depend in a pointwise way on the source terms, it seems that
one needs to control derivatives of these source terms.
In the corresponding results for the Vlasov-Poisson or
Vlasov-Maxwell systems one exploits the fact that the field 
quantities depend on the source terms through spatial or spacetime
integrals; the corresponding field equations are smoothing,
cf.\ \cite{Batt77,GlStr1,Rein07}. Here the field equation (\ref{ein4})
provides this smoothing effect, because it turns out that
in order to control the derivatives of the characteristic flow with
respect to the initial data only a certain combination of
second order derivatives of the metric coefficients is needed,
and this combination is precisely the one which appears 
in (\ref{ein4}) and is therefore controlled by $q$.

In the context of these arguments and also in what follows 
some further information is useful.
By (\ref{ein1}) and (\ref{ein2}), 
\[
\partial_r\mu+\partial_r\lambda\geq 0, 
\]
and together with the boundary condition
(\ref{boundcinf}) and (\ref{e2lamb}) this implies that
\be \label{mulambest}
\mu -\lambda \leq \mu +\lambda \leq 0. 
\ee
Solutions of the Einstein-Vlasov system satisfy the following conservation
laws. The Vlasov equation implies that the number of particles
\be \label{partn}
N = \iint e^{\lambda (t,r)}  f(t,x,v)\, dv\,dx
\ee 
is conserved. More importantly, if we observe that $j$ is for each fixed
time $t$ compactly supported so that by (\ref{ein3}), $\partial_t\lambda(t,r)=0$
for $r$ sufficiently large, we conclude from (\ref{e2lamb}) that
the ADM mass
\be \label{adm}
M = \lim_{r\to \infty}m(t,r) = \iint \sqrt{1+|v|^2} f (t,x,v) \, dv\, dx 
\ee
is conserved.

\subsection{Global existence for small data}

A natural question when investigating a nonlinear system of PDEs is whether
small initial data lead to global solutions which disperse. As mentioned
above, no such smallness condition can be formulated for the at least
formally closely related Einstein-dust system, cf.\ \cite{Chr84},
but for true Vlasov matter such a global result for small data does exist. 
\begin{theorem} \label{glex}
For all $R >0$ there exists $\epsilon >0$ such that
if $f$ is a maximal solution of the asymptotically flat, spherically 
symmetric Einstein-Vlasov system with $f(0)=\fn$ satisfying
\[
\fn\,(x,v) = 0 \ \mbox{for}\ |x| + |v| > R 
\]
and
\[
||\fn\,||_\infty  < \epsilon,
\]
then the solution exists globally in $t$. Moreover, the solution disperses
in the sense that
\[
||\rho (t)||_\infty \leq C (1 +|t|)^{-3},\ t\in \R, 
\]
and the spacetime is geodesically complete.
\end{theorem}
This result was first proven in \cite{RR1}, see also \cite{Rein95,Rend97}.
Similar results were known both for the Vlasov-Poisson and the
Vlasov-Maxwell system, cf.\ \cite{BD,GlStr2,Rein07}.
The basic dispersive mechanism in the system can be seen as follows.
By (\ref{fviachar}) and the change of variables formula,
\beas
\rho(t,x) 
&=& 
\int \sqrt{1+|v|^2} \fn\, ((X,V)(0,t,x,v))\, dv \\
&=& 
\int \sqrt{1+|v|^2} \fn\, (X,V(0,t,x,v))\, 
\left|\det \left(\frac{\partial X(0,t,x,v)}{\partial v}\right) \right|^{-1}dX.
\eeas
Now in the flat Minkowski case, $\lambda = \mu = 0$,
$X(0,t,x,v) = x - t \,v /\sqrt{1+|v|^2}$, and the determinant above
grows like $t^3$. And if the field terms satisfy suitable decay
conditions this turns out to remain correct. The decay of $\rho$ and
the other source terms in turn implies decay for the field terms, and
a bootstrap argument gives the result. Here it again becomes important
that a certain combination of second order derivatives of the metric 
coefficients can be controlled via the source terms by the field equation
(\ref{ein4}).

The above argument proves that the solution is global with respect to
the chosen time coordinate. However, this does not automatically imply
that the maximal Cauchy development of the corresponding initial data
is singularity-free. By definition, a spacetime contains a singularity
if there exists a timelike or null geodesic, i.e., a world line
of a particle or a photon, whose maximally extended domain of affine
parameter, i.e., proper time in case of a particle, is not the
whole real line. The geometrically invariant, coordinate-free
characterization of a singularity-free or global spacetime
is therefore that all timelike and null geodesics exist on
the whole real line. In the course of the proof of the above bootstrap
argument one obtains sufficient decay information on the Christoffel
symbols to conclude that indeed all maximally extended geodesics
in this spacetime are complete, i.e, exist on the whole real line.

A mechanism which also leads to global existence but is different from
dispersion induced by small data is that initially all the particles 
move outward sufficiently fast to prevent re-collapse, cf.\ \cite{AKR1}.
For a collisionless gas of massless particles a result analogous
to Theorem~\ref{glex} has been shown in \cite{D06}.

\subsection{Possible break down must occur at the center first}

The next question is whether large data lead to a break
down of the solution. It is important to realize 
that the fact that---as shown in the next section---certain data do 
lead to gravitational collapse and the formation of black holes does
certainly not rule out the possibility that in Schwarzschild
coordinates all solutions of the spherically symmetric, asymptotically
flat Einstein-Vlasov system are global. 
If the weak cosmic censorship conjecture
holds for this system, then, physically speaking, 
an observer who is very far away (ideally at spatial infinity) should
not be able to observe the singularity
so that for him the universe should look singularity-free,
and in this context one should note that Schwarzschild time
asymptotically coincides with the proper time measured by such 
an observer.
So far, all numerical simulations and analytical results support 
the conjecture that all solutions are global in Schwarzschild time. 
In all cases where
a gravitational collapse was simulated numerically the solution
did not seem to break down in finite Schwarzschild time and an
event horizon did form, cf.\ \cite{AR1, OC, RRS2}.
In the next section we discuss a recent analytical result
which shows that at least for a certain class of data solutions
do again exist globally in Schwarzschild time, but they do undergo
a gravitational collapse and form a black hole. In the context of
that result it is important to know that if a solution should
develop a singularity in finite Schwarzschild time, then this must
happen at the center first.
\begin{theorem} \label{reg}
Let $f$ be a regular solution of the spherically symmetric,
asymptotically flat Einstein-Vlasov system
on a time interval $[0,T[$. 
Suppose that there exists an open neighborhood $U$ 
of the point $(T,0)$ such that
\be \label{vbound}
\sup\{|v| \mid (t,x,v)\in\supp f\cap (U\times\R^3)\}<\infty.
\ee
Then $f$ extends to a regular solution on $[0,T'[$
for some $T'>T$.
\end{theorem}
This theorem was established in \cite{RRS1}. In order to explain the mechanism
behind this we introduce coordinates on the mass shell which are adapted to
the symmetry:
\[
r = |x|,\ w = \frac{x\cdot v}{r},\ L = |x\times v|^2.
\]
Spherical symmetry of $f$ implies that by abuse of notation,
\[
f(t,x,v) = f(t,r,w,L).
\]
One should think of $w$ as the non-canonical radial momentum and
$L$ as the modulus of angular momentum squared of a particle.
Due to spherical symmetry $L$ is conserved along characteristics,
and the characteristic system written in $(r,w,L)$ takes the form
\begin{eqnarray}
\dot r
&=&
e^{\mu - \lambda} \frac{w}{E}, \label{rdot}\\
\dot w
&=&
- \left( \partial_t\lambda\, w + e^{\mu - \lambda} \partial_r\mu\, E 
- e^{\mu - \lambda} \frac{L}{r^3 E} \right),\label{wdot}\\
\dot L
&=&
0, \label{Ldot}
\end{eqnarray}
where
\[
E=E(r,w,L):=\sqrt{1+w^{2}+L/r^{2}} = e^\mu p^0.
\]
The source terms take the form
\begin{eqnarray}
\rho(t,r)
&=&
\frac{\pi}{r^{2}}
\int_{-\infty}^{\infty}\int_{0}^{\infty}Ef(t,r,w,L)\,dL\,dw,
\label{rhorwl}\\
p(t,r)
&=&
\frac{\pi}{r^{2}}\int_{-\infty}^{\infty}\int_{0}^{\infty}
\frac{w^{2}}{E}f(t,r,w,L)\,dL\,dw,
\label{prwl}\\
j(t,r)
&=&
\frac{\pi}{r^{2}}
\int_{-\infty}^{\infty}\int_{0}^{\infty}w\,f(t,r,w,L)\,dL\,dw;
\label{jrwl}
\end{eqnarray}
the quantity $q$ is not needed in what follows.
Since on the support of $f$ the quantity $L$ is bounded initially,
it remains bounded for all time. In order to establish the theorem
one needs to control the increase in $|v|$ along characteristics
which stay away from the center. Because of the relation 
$|v|^2 = w^2 + L/r^2$ it is sufficient to control $w$ 
along such a characteristic. If we substitute the expressions
for $\partial_t\lambda$ and $\partial_r\mu$ and for the source terms 
into (\ref{wdot}), we find that
\beas
\dot w
&=& 
e^{\mu-\lambda} \frac{L}{r^3 E} - e^{\mu + \lambda} \frac{m(s,r)}{r^2} E \\
&&
{}+  4\pi r e^{\mu +\lambda }\frac{\pi}{r^2} \int_{-\infty}^{\infty}\int_{0}^{\infty}
\left(w \tilde w - E \frac{\tilde w^2}{\tilde E}\right)\, 
f(s,r,\tilde w,\tilde L)\,d\tilde L\,d\tilde w\ \ 
\eeas
with obvious definition of $\tilde E$. Now $L$ is bounded, $r$ is bounded
away from zero, $m$ is bounded by $M$, and 
$\mu-\lambda \leq \mu + \lambda \leq 0$ by (\ref{mulambest}). 
Hence the term to focus on is the double integral. But
\beas
w \tilde w - E \frac{\tilde w^2}{\tilde E}
&=&
\frac{\tilde w}{\tilde E}
\frac{w^2 (1+ \tilde w^2 + \tilde L/r^2)-\tilde w^2 (1+w^2 + L/r^2)}
{w \tilde E + \tilde w E} \\
&=& 
\frac{\tilde w}{\tilde E}
\frac{w^2 (1+\tilde L /r^2) - \tilde w^2 (1+ L/r^2)}
{w \tilde E + \tilde w E}, 
\eeas
provided the denominator does not vanish.
In the last step the term $w^2 \tilde w^2$, which is the worst
one concerning powers of $w$ or $\tilde w$, canceled, and this is the
crucial observation in order to establish Theorem~\ref{reg},
since it leads to an estimate of the form
\[
P(t) \leq P(0) + C \int_0^t P(s)\, \ln P(s)\, ds,
\]
cf.\ the continuation criterion in Theorem~\ref{le}.

The result above refers to the Einstein-Vlasov system written in Schwarzschild
coordinates. The analogous result in so-called maximal-isotropic coordinates
was established in \cite{Rend97}. In \cite{DR05} the analogous result is 
established using coordinates which cover the whole spacetime, even in 
the presence of trapped surfaces.

To appreciate this result it is instructive to think of a spherically
symmetric solution of the Einstein-dust system. In that system all particles
move radially, and particles at the same radius $r$ have the same momentum,
i.e., they remain on the same sphere which can grow or shrink as a whole, 
cf.\ (\ref{dustf}).
If two such spheres cross, the system experiences a so-called shell-crossing
singularity which occurs at some positive radius. If one wants to
study the formation of event horizons and spacetime singularities in
the Einstein-dust system, one has to handle the problem that the solution
may break down due to a shell-crossing singularity before the objects
of interest have actually evolved.

\section{Gravitational collapse and formation of black holes} \label{seccollapse}
\setcounter{equation}{0}
As pointed out repeatedly above, Schwarzschild coordinates do not
cover regions of spacetime which contain trapped surfaces.
Since in gravitational collapse the latter typically appear
before a singularity forms, one may well argue that in order to
investigate gravitational collapse and the formation of black holes,
one should better not use these coordinates
to begin with. On the other hand, they do have advantages from an
analysis point of view, and one may still hope to derive the desired
information on gravitational collapse from the asymptotic behavior
of the solution for large Schwarzschild time. But for large data which 
possibly lead to gravitational collapse there is so far no global existence result
in Schwarzschild time. In order to by-pass this difficulty we study
the solutions in a coordinate region which avoids the center,
on which we can consistently formulate and study the corresponding
Cauchy problem, and which is large enough to allow us to conclude the
formation of black holes. That the behavior of the solution
on this coordinate region indeed implies the weak cosmic censorship
conjecture and the formation of
a black hole is shown in Section~\ref{subsecbh},
where we rely on a mathematical formulation of the corresponding concepts
given in \cite{Chr99}. 

\subsection{The system on an outer domain}

We study the solutions to the spherically symmetric, asymptotically flat 
Einstein-Vlasov system 
({\ref{vlasov})--(\ref{q}) on the exterior region
\begin{equation} \label{ddef}
D:=\{(t,r) \in [0,\infty[^2 \mid r \geq \gamma^+(t)\},
\end{equation}
where $\gamma^+$ is an outgoing radial null geodesic
originating from some $r=r_0>0$, i.e.,
\begin{equation}\label{gamma+}
\frac{d \gamma^+}{ds}(s)=e^{(\mu-\lambda)(s,\gamma^+(s))},\;\gamma^+(0)=r_0.
\end{equation}
This equation is obtained from (\ref{rdot}) by replacing the $1$ in the
definition of $E$ by $0$, i.e., the particle under consideration
has rest mass $0$ like a photon should, and by putting $L=0$, i.e.,
the photon moves radially. That the photon is outgoing, i.e., 
moving away from the center, means that $w>0$, and hence $w/E = 1$.
In order to restrict the analysis of the system to the region $D$
we have to find a replacement for the boundary 
condition (\ref{boundc0}). We prescribe the total ADM mass
$M>0$ and redefine 
the quasi-local mass by
\begin{equation}\label{m-def}
m(t,r)=M - 4\pi\int_r^\infty \rho(t,\eta)\,\eta^2 d\eta,
\end{equation}
while retaining the definition (\ref{e2lamb}) for $\lambda$. Clearly,
a solution of the system as considered in the previous 
section, when restricted to $D$, is a solution of this modified system.
Moreover, characteristics of the Vlasov equation can pass
from the region $D$ into the region $\{r < \gamma^+(t)\}$
but not the other way around so that initial data
$\fn$ posed for $r>r_0$ completely determine the solution
on the outer domain $D$. Such data need to satisfy the restrictions
(\ref{init2mor})
and 
\begin{equation}\label{mout}
M_\mathrm{out} = 4 \pi \int_{r_0}^\infty \open{\rho}(\eta)\,\eta^2 d\eta < M,
\end{equation}
where $\open{\rho}$ is induced by $\fn$.
Then $\lim_{r\to\infty} m(t, r)=M$ and $0\le m\le M$.
The crucial question is whether one can specify data such that
$\gamma^+$ has the property that
\begin{equation}\label{limgamma}
   \lim_{s\to \infty}\gamma^+(s) < \infty.
\end{equation}
While this is not sufficient to conclude the formation of a black
hole, it turns out to be the main step towards that goal.

It is important to note that the behavior or even the nature of
the matter in the region $\{r < \gamma^+(t)\}$ is not going to
be relevant in what follows. For example one can equally well
think of Vlasov data being posed on $\{ r \geq 0\}$, but only the data
on $\{ r \geq r_0\}$ need to be properly restricted in order to
obtain the desired behavior of the solution on the outer domain $D$. 
What is essential is that there initially
is and hence remains some matter in the region $\{r < \gamma^+(t)\}$
as guaranteed by the condition $M_\mathrm{out} < M$. 

\subsection{The main result---analysis in Schwarzschild coordinates}

The initial data 
$\fn \in C^1_c(\R^6)$ for the outer matter should satisfy the
condition that on $\supp \fn$,
\[
R_0 \leq r \leq R_1,\ w \leq W_-
\]
where we use the variables $(r,w,L)$ introduced above, and
\[
0 < r_0 < R_0 < R_1,\ W_- < 0.
\]
In particular, all particles move inward initially. These parameters
can be specified in such a way that the data are close to violating the
condition $2 \open{m}(r)/r < 1$, that the particles continue to move inward 
on the outer domain $D$, and that (\ref{limgamma}) holds.
The main result is the following. 
\begin{theorem} \label{bh}
There exists a class of regular initial data for
the spherically symmetric Einstein-Vlasov system such that 
for such data the corresponding solution exists on $D$,
and
\[
\lim_{s\to \infty}\gamma^+(s) < \infty.
\]
In addition the following holds:
\begin{itemize}
\item[(a)]
For $t\to \infty$ no matter remains in the region $\{ r> 2 M\}$, more
precisely,
the solution is vacuum and the metric equals
the Schwarzschild metric (\ref{scm}) with mass $M$ in the region
\[
t \geq 0 \ \mbox{and}\ r \geq 2 M + \alpha e^{-\beta t}.
\] 
Here $\alpha, \beta >0$ depend only on the initial
data class.   
\item[(b)]
In the outer region $D$ and for $r \leq 2 M$,
\[
\lim_{t\to \infty} \mu(t,r) = -\infty,
\]
more precisely, 
\[
\mu(t,r) 
\leq \ln \left(\frac{\alpha e^{-\beta t}}{2 M + \alpha e^{-\beta t}}\right)^{1/2} 
\]
for all $t\geq 0$ and $\gamma^+ (t) \leq r \leq 2 M + \alpha e^{-\beta t}$.
This implies that for $c\leq 2M$ the timelike lines $r=c$  are incomplete,
i.e., they have finite proper length, and this length is uniformly bounded.
\item[(c)]
The radially outgoing null geodesic which does not escape
to infinity and is furthest to the right with this property
gets trapped precisely at the Schwarzschild radius $r=2 M$. 
More precisely, we define
\beas
r^\ast := \sup \{ r\geq r_0
&|&
\mbox{the radially outgoing null geodesic}\ \gamma\ \mbox{with}\\
&&
\gamma(0)=r\ \mbox{satisfies}\ 
\lim_{s\to\infty}\gamma (s) < \infty \},
\eeas
and let $\gamma^\ast$ be the radially outgoing null geodesic
with $\gamma^\ast (0) = r^\ast$. Then
\[
\lim_{s\to\infty}\gamma^\ast (s) = 2 M,
\]
and every  radially outgoing null geodesic $\gamma$ with $\gamma(0) > r^\ast$
is future complete, i.e., it exists on $[0,\infty[$ in an affine
parameterization, and $\lim_{s\to\infty}\gamma (s) = \infty$.
\end{itemize}
\end{theorem}
If a solution with mass $M$ of the spherically symmetric, asymptotically flat
Einstein-Vlasov system collapses to a black hole of mass $M$
and if we coordinatize this solution by Schwarzschild coordinates, then
the asymptotic behavior as $t\to \infty$ of this solution should
be as given by this theorem. In the next subsection
we show that this behavior actually implies that a black hole forms
in the sense of a suitable, coordinate-independent formulation
of this concept.

For a complete proof of Theorem~\ref{bh} we refer to \cite{AKR2}.
Here we want to highlight some central arguments. 
The first of these makes sure that the particles in the outer 
domain $D$ keep moving inward in a controlled way.
Since initially $w \leq W_- < 0$ for all particles,
this remains true on some time interval.
On this time interval and along any characteristic in $\supp f$,
\begin{eqnarray*}
\frac{d}{ds}(e^{-\lambda}w) 
& = &
- \frac{4\pi^2}{r}\,e^{\mu}\,\int_{-\infty}^\infty\int_0^\infty
\bigg[\sqrt{\frac{\tilde{E}}{E}}\,w
-\sqrt{\frac{E}{\tilde{E}}}\,\tilde{w}\bigg]^2\,f\,d\tilde{L}\,d\tilde{w}\\ 
&&
{}-e^{\mu}\frac{m}{r^2}\bigg(\frac{1+L/r^2}{E}+\frac{2L}{r^2 E}\bigg)
   +e^{\mu}\frac{L}{r^3 E}\\
&\le&
 -e^{\mu}\frac{m}{r^2}\bigg(\frac{1+L/r^2}{E}
   +\frac{2L}{r^2 E}\bigg)+e^{\mu}\frac{L}{r^3 E}.
\end{eqnarray*}
Differentiating (\ref{e2lamb}) w.r.t.\ $t$ and using (\ref{ein3})
leads to $\partial_t m =-4\pi r^2 e^{\mu-\lambda} j \geq 0$
on the time interval we consider.
It follows that $m(s, r)\ge m(0, r)=\open{m}\,(r)$.
Thus as long as the characteristic remains in $D$,
\[
\frac{d}{ds}(e^{-\lambda}w)
\le 
e^{\mu}\,\frac{1}{r^3 E}\bigg(L-\frac{3L}{r}\,\open{m}\,(r)
-r\,\open{m}\,(r)\bigg).
\]
We require that 
\begin{equation}\label{hypoL}
   0< L <\frac{3L}{r}\,\open{m}(r) + r\,\open{m}(r),\ r\in [r_0,R_1]
\end{equation}
on $\supp \fn$. Then the above estimate together with a 
bootstrap argument and the fact that $e^\lambda \geq 1$ show that 
\begin{equation}\label{west}
w \le\Big(\min_{r\in [R_0, R_1]} e^{-\lambda(0,r)}\Big)\,W_-
\end{equation}
on $\supp f \cap D$.

Before going further some comments on the condition
(\ref{hypoL}) are in order. By our general set-up some
mass must be inside $\{ r\leq r_0\}$ initially, and
this mass is a lower bound on $\open{m}\,$ in (\ref{hypoL}).
The role of this mass is not to pull the particles inward, but to
keep them focused towards the center. The smaller their
angular momentum is, i.e., the better they are aimed straight towards
the center, the smaller can the central mass be chosen initially.
Notice that for spherically symmetric dust which is used as
matter model in \cite{OS} and \cite{Chr84} all particles have
angular momentum equal to zero.

The second issue in the proof of Theorem~\ref{bh} we want to touch upon
is the limiting behavior of $\gamma^+$. The basic idea is to consider
a radially ingoing null geodesic $\gamma^-$ which starts to the left of
the outer matter and to the right of $\gamma^+$, i.e., 
\[
\frac{d\gamma^-}{ds}(s)= - e^{(\mu-\lambda)(s,\,\gamma^-(s))},
   \quad r_0<r_1=\gamma^-(0)< R_0.
\]
Then initially and therefore as long as $\gamma^+$ and  $\gamma^-$ 
do not intersect there is no matter in the region
between the outgoing and the ingoing null geodesic. This fact 
can be used to estimate their relative velocity in such a way
that in Schwarzschild time they actually never intersect.
This proves the limiting behavior of $\gamma^+$ and 
furthermore shows that the matter which is initially in $D$ 
stays strictly to the right of $\gamma^+$ and therefore in $D$
for all future Schwarzschild time. In order to estimate how far
the two null geodesics can move at most we observe that
by (\ref{mu}),
\[
\mu (t,r) \leq -\int_r^\infty e^{2\lambda(t,s)} \frac{m(t,s)}{s^2} ds 
=: \hat \mu (t,r),
\]
and hence
\[
|\dot \gamma^\pm| \leq e^{\hat \mu}.
\]
We wish to see that the right hand side becomes small,
and to this end we observe that the following chain of estimates
yields a lower bound for $\partial_t \hat\mu$:
\begin{eqnarray}
1-e^{(\mu+\lambda)(t,r)} 
&=& 
\int_r^\infty (\partial_r\lambda + \partial_r\mu)(t,\eta)\, e^{\mu+\lambda}
d\eta
\nonumber \\ 
&=&
4\pi \int_r^\infty \eta\,(\rho+p)(t,\eta)
e^{\mu+\lambda}e^{2\lambda}\,d\eta\nonumber\\
&\leq& 
4\pi \int_r^\infty 3\,\eta  \, (- j (t,\eta))
\,e^{\mu+\lambda}e^{2\lambda}\,d\eta\nonumber\\
&=&
- 3 \int_r^\infty \eta \frac{\partial}{\partial t}
\left(e^{2\lambda}\frac{m(t,\eta)}{\eta^2}\right) \,d\eta \nonumber\\
&\leq& 
-3 R_1 \partial_t\hat\mu (t, r). \label{dthatmu}
\end{eqnarray}
In the first estimate we exploit the fact that not only is
$j$ negative, but by choosing $|W_-|$ large the
source terms $\rho$ and $p$ can be estimated by a suitable
multiple of $-j$. 

If one puts all the details which are left out here together,
one obtains a list of conditions on the initial data which
make sure that these estimates hold true for all future
Schwarzschild time. The resulting conditions are the following:
\[
\mathrm{supp}\, \fn \subset [R_0, R_1]\times ]-\infty,W_0] \times [0,\infty[
\] 
with
\[
0 < r_0 < r_1 = 2 M < R_0 = \frac{r_1 + R_1}{2} < R_1,\ W_- < 0,
\] 
\begin{equation} \label{m0cond}
0 < \open{m}(r_0) < M, \quad \frac{2 \open{m}(r_0)}{r_0} < \frac{8}{9}
\end{equation}
and either
\[
R_1-r_1<\frac{r_1-r_0}{6},
\]
or
\[
\sqrt{1-\frac{r_1}{R_1}}
< \min\left\{\frac{1}{6},\frac{r_0^2}{36 R_1 M},
\frac{r_1-r_0}{24 R_1}\right\}.
\]
Since $r_1 = 2 M$, the latter two conditions both say
that $2 M / R_1$ must be close to $1$. Once $r_0,r_1,R_0,R_1$, 
and $\open{m}(r_0)$ have been chosen, $|W_-|$ has to be chosen
sufficiently large where we refrain from making this precise here. 
Any initial distribution
on $\{r>r_0\}$ which satisfies (\ref{init2mor}) and (\ref{hypoL})
is admissible in Theorem~\ref{bh}, and it is easy to see that there
exist such data.

Since the various parameters which enter the definition of our class
of admissible data are defined in terms of inequalities,
the set of data has ``non-empty interior'',
in the sense that sufficiently small perturbations
of initial data in the ``interior'' of this set belong to it as well.

The crucial step in the proof of the remainder of Theorem~\ref{bh}
is to show that all particles move towards $r=2 M$ with the stated estimate.
As a first step note that by (\ref{e2lamb}) and (\ref{mu}),
\[
(\mu-\lambda)(t,r)
\geq
\ln\frac{r-2M}{r},\ r> 2 M.
\]
Together with the control for the radial momentum of the particles
this implies that along any characteristic in the matter support,
\[
\dot r = e^{(\mu-\lambda)(s,r)}\frac{w}{E} \leq - C e^{(\mu-\lambda)(s,r)}
\leq - C \frac{r-2M}{r}
\]
as long as $r > 2 M$; $C>0$ is determined by the initial data parameters.  
Integrating this differential inequality proves the support estimate in 
the theorem.

The spherically symmetric, asymptotically flat Einstein-Vlasov system has a
wide variety of static solutions with
finite ADM mass and compact matter support,
cf.\ \cite{Rss1, RRss1, RRss2}. Particularly interesting examples of initial
data to which our results apply are obtained if the matter for $r \leq r_0$
is represented by such a static solution.

\begin{cor} \label{ssinthemiddle}
Let $f_s$ be a static solution of the spherically symmetric,
asymptotically flat Einstein-Vlasov
system with finite ADM mass $M_s>0$ and finite radius $r_s >0$. Define
$r_0=r_s$, let $r_1 > r_0$ be arbitrary, $M=r_1/2$, and 
$M_\mathrm{out} = M -M_s$; the latter quantity is positive. 
Then one can construct data $\open{f}$ on $\{ r\geq r_s\}$
such that the solution of the Einstein-Vlasov system launched by $f_s + \fn$
has the properties stated in Theorem~\ref{bh}, 
it exists for all $t\geq 0$ and $r\geq 0$,
and it coincides with the static solution $f_s$
for all $r\leq \gamma^+(t)$ and $t\geq 0$.
\end{cor}
It is at this point that the choice $8/9$ in (\ref{m0cond}) is relevant;
for the proof of Theorem~\ref{bh} any positive constant less than $1$ 
would do. But in \cite{And2} it is shown that for any steady
state of the spherically symmetric Einstein-Vlasov system
the condition $2 m(r)/r < 8/9$ holds for all $r>0$,
and this bound is actually sharp, cf.\ \cite{And3}.

Given the fact that the solutions described in Theorem~\ref{bh}
undergo gravitational collapse and form a black hole it may seem
strange that in the center of such a solution a steady state 
comfortably sits for all $t\geq 0$. But this is of course due to
the fact that $t$ is Schwarzschild time. In the region $r < 2 M$
the solution can be extended using a different time coordinate,
and if the latter is properly chosen then the outer matter will
all pass within $r < 2 M$, it will crash into the steady state, and
all the matter will finally collapse into a spacetime singularity.
To support all phases of this evolution with rigorous theorems
is one of our projects for current and future research, cf.\
Section~\ref{secpers}.

On the other hand, the fact that the steady state sits undisturbed
in the center as long as none of the outer matter reaches it is easily
understood from a physics point of view, and just as easily proven.
In spherical symmetry particle orbits within a certain radius $r$
are not 
influenced by matter which is at strictly larger radii. 
Hence adding the outer matter shell does not change the fact that 
the steady state satisfies the Einstein-Vlasov system on $\{r \leq r_0\}$
as long as the outer matter remains outside that region. 

\subsection{Weak cosmic censorship and the formation of a black hole} \label{subsecbh}

The result of the previous section shows in particular
that no particle and no light ray can escape from the
region $\{ r \leq \gamma^\ast (t)\}\subset \{r \leq 2 M\}$. Since 
outside of this domain the
solution is global, this means in particular that no
causal curve originating in a possible singularity
can reach the region $\{r > 2 M\}$. However, all this
refers to the formulation of the Einstein-Vlasov system
in Schwarzschild coordinates, and we are not (yet) allowed
to conclude that our spacetime satisfies the weak cosmic censorship 
conjecture and that a black hole forms. In the present subsection
we address these questions in a coordinate-independent manner.
We start with showing that spacetimes as obtained in the previous
subsection satisfy the weak cosmic censorship conjecture which in heuristic
terms says that no singularities can be observed from infinity even if
the observations are allowed to continue indefinitely.
Expressed more precisely we have to show:

\begin{proposition} \label{wcc}
The spacetimes obtained in Theorem~\ref{bh} possess a complete
future null infinity.
\end{proposition} 
The concept of future null infinity usually refers to a conformal
compactification of the spacetime under investigation which attaches
a boundary at infinity, cf.\ \cite[11.1]{Wald}. We prefer to follow
Christodoulou \cite[p.~A26]{Chr99} for the definition of the term
``possess a complete future null infinity''. 

The set $B_0 :=\{ (0,r) \mid r < R_1\}$ has compact closure,
and the boundary $C_0^+$ of its causal future is given by the
radially outgoing null geodesic $\gamma_1$ starting at $R_1$;
we recall that $R_1$ is the outer radius of the initial matter
support in Theorem~\ref{bh}.
By that theorem, $\gamma_1$ is future complete. Consider now
a domain $B:=\{ (0,r) \mid r < R_2\}$ with $R_2 > R_1$ and
the boundary $C^-$ of its domain of dependence which is given
by the radially ingoing null geodesic $\gamma_2$ starting at $R_2$.
According to the definition in \cite{Chr99} we must show that
the affine length of $\gamma_2$, measured from the intersection
of $\gamma_1$ and $\gamma_2$, goes to infinity as $R_2 \to \infty$.
In doing so the affine parameterization of  $\gamma_2$ must
be normalized in such a way that its tangent at $(0,R_2)$ equals
the vector $T-N$ where $T$ is the future directed unit normal
to the initial hypersurface $\{ t=0 \}$, i.e., $T=(e^{-\mu(0,R_2)},0,0,0)$,
and $N$ is the outward unit normal to $B$ in the initial hypersurface,
i.e., $N=(0,e^{-\lambda(0,R_2)},0,0)$. 

It turns out that already the affine length of $\gamma_2$ between
the intersections with $\gamma_1$ and with the line $r=R_1$ goes to
infinity as $R_2 \to \infty$. But on the region $\{r\geq R_1\}$
we have according to Theorem~\ref{bh} vacuum with metric given
by
\[
e^{2 \mu (t,r)} = 1-\frac{2 M}{r} = e^{-2 \lambda (t,r)}.
\]
When parameterized by coordinate time radial null geodesics
in the region $\{r\geq R_1\}$ satisfy the estimates
\[
1-\frac{2 M}{R_1} \leq 1-\frac{2 M}{r} = 
e^{\mu - \lambda} = \left|\frac{dr}{dt}\right| \leq 1.
\]
Let $(T^\ast,R^\ast)$ denote the point where $\gamma_1$ and $\gamma_2$
intersect. By the above estimate, $\gamma_1(t) \leq R_1 + t$ and  
$\gamma_2(t) \geq R_2 - t$ which implies that $T^\ast \geq (R_2 - R_1)/2$
and hence
\[
R^\ast \geq R_1 + \frac{R_2 - R_1}{2} \left(1-\frac{2 M}{R_1}\right).
\]
Consider now an affine parameterization $\tau \mapsto (t,r,\theta,\varphi)(\tau)$
of $\gamma_2$, $\tau \geq 0$, with $(t,r)(0)=(0,R_2)$. 
Since $\gamma_2$ is radial, $\theta=\pi/2,\
\varphi=0$. Since $\gamma_2$ is null, 
\[
-e^{2\mu}\left(\frac{dt}{d\tau}\right)^2 + e^{2\lambda}
 \left(\frac{dr}{d\tau}\right)^2 =0,
\]
and since $\gamma_2$ is ingoing, i.e., $dr/d\tau< 0$, we find that
\[
\frac{dt}{d\tau} =- e^{\lambda-\mu} \frac{dr}{d\tau}.
\]
By the geodesic equation, 
\begin{eqnarray}
\frac{d^2 r}{d\tau^2} 
&=&
 -e^{2(\mu - \lambda)} \partial_r\mu \left(\frac{dt}{d\tau}\right)^2 - 
\partial_r\lambda \left(\frac{dr}{d\tau}\right)^2
-2 \partial_t\lambda \frac{dt}{d\tau} \frac{dr}{d\tau}\nonumber \\
&=&
4\pi r e^{2\lambda}\left(\frac{dr}{d\tau}\right)^2 [-p-\rho-2j] = 0,
\label{rddotinvac}
\end{eqnarray}
i.e., $dr/d\tau = const =: \sigma$ as long as $\gamma_2$ is in the 
vacuum region $\{r\geq R_1\}$. 
The normalization condition mentioned above requires that
\[
\left(\frac{dt}{d\tau},\frac{dr}{d\tau}\right)(0)
= \left(-e^{\lambda(0,R_2) - \mu(0,R_2)}\sigma,\sigma\right) =
\left(e^{-\mu(0,R_2)},-e^{-\lambda(0,R_2)}\right)
\]
which means that we should choose $\sigma :=-e^{-\lambda(0,R_2)}$.
Let $\tau_1$ and $\tau_2$ be the values of the affine parameter such
that $(t,r)(\tau_1)$ is the intersection point
of $\gamma_2$ with $\gamma_1$ and $(t,r)(\tau_2)$ is 
the intersection point of $\gamma_2$ with the
line $r=R_1$, i.e., $r(\tau_1)=R^\ast$ and $r(\tau_2)=R_1$. For the affine length $L$ of 
the corresponding piece of $\gamma_2$ we find that
\beas
L 
&=&
\int_{\tau_1}^{\tau_2} d\tau = \int_{R^\ast}^{R_1} \frac{dr}{\sigma}
= e^{\lambda(0,R_2)} (R^\ast - R_1)\\
&\geq&
\left(1-\frac{2 M}{R_2}\right)^{-1/2}
\frac{R_2-R_1}{2} \, \left(1-\frac{2 M}{R_1}\right) \\
&\to&
\infty\ \mbox{as}\ R_2 \to \infty.
\eeas
This proves the claim of Proposition~\ref{wcc} in the sense of \cite{Chr99}.

We now turn to the question whether Theorem~\ref{bh} also implies the
formation of a black hole in some appropriate, coordinate-free sense.
A maximal development of Cauchy data is said to contain a black hole
if future null infinity $\mathcal{I}^+$ is complete and the causal past
$J^-(\mathcal{I}^+)$ of future null infinity has non-empty complement,
cf.\ \cite[Sect.~12]{DR05}. Intuitively this says that no causal
curve, i.e., no particle trajectory or light ray, originating
in the complement of $J^-(\mathcal{I}^+)$ can reach future null
infinity, and that such trapped causal curves really exist. 
In the spacetimes obtained in Theorem~\ref{bh} the null geodesic
$\gamma^\ast$ does not reach future null infinity, and since
we have shown that the latter is actually complete, $\gamma^\ast$
cannot reach future null infinity in any extension (such as 
the maximal development) either. Since $\mathcal{I}^+$ is complete
for the spacetimes we obtained, $\gamma^\ast$, in order to reach 
$\mathcal{I}^+$ in the maximal development, would have to enter
the outer region $D$; $\gamma^\ast(\tau^\ast) \in D$. But following
$\gamma^\ast$ 
backwards we see that 
it must have stayed in $D$ for $\tau \leq \tau^\ast$ and hence cannot have
reached the region $\{ r > 2 M\}$ to begin with. 

In other words, we have in Theorem~\ref{bh} constructed a globally
hyperbolic spacetime $\mathcal{M}$ which has a complete $\mathcal{I}^+$
and in which the complement of $J^-(\mathcal{I}^+)$ is non-empty.
If we consider the maximal development $\overline{\mathcal{M}}$
of the same Cauchy data, it has the same, complete $\mathcal{I}^+$, 
and since $\mathcal{M}\subset\overline{\mathcal{M}}$, the
complement of $J^-(\mathcal{I}^+)$ in $\overline{\mathcal{M}}$
is non-empty as well.
Hence the following is true.  

\begin{proposition} \label{nomoredoubts}
Initial data as specified in  Theorem~\ref{bh}
lead to the formation of a black hole in the following sense.
The spacetimes $\mathcal{M}$ obtained in Theorem~\ref{bh} possess a complete
future null infinity $\mathcal{I}^+$, and its causal past
$J^-(\mathcal{I}^+)$ has non-empty complement
in the maximal development of the data.
\end{proposition}

\section{Concluding remarks}
\setcounter{equation}{0}

In this final section we want to discuss the question to which 
extent the results from the previous section
really depend on the particular matter 
model which we employed, we want to compare these results 
to previous results for other matter models, and we want to
discuss future perspectives.  
\subsection{General matter models} \label{secgenmat}
The issue of gravitational collapse and in particular
the validity of the cosmic censorship conjecture should of 
course be addressed not only for one particular matter model
like the collisionless gas. Indeed, some key arguments in the proof
of Theorem~\ref{bh} do depend only on certain general properties
of the matter model and not on its specific nature.
We briefly discuss this issue.
To this end, let
\begin{equation} \label{genmatquant}
\rho := e^{-2\mu} T_{00},\ p := e^{-2\lambda} T_{11},\
j := - e^{-\mu-\lambda}T_{01}.
\end{equation}
Firstly, we  assume that the following two conditions
are satisfied.
\begin{itemize}
\item The dominant energy condition holds. \hfill(DEC)
\item The radial pressure $p$ is non-negative. \hfill(NNP)
\end{itemize}
In general relativity the dominant energy condition (DEC) is the 
main criterion which a matter model must satisfy in order
to be considered realistic, cf.\  \cite{HE}. 
The non-negative pressure condition (NNP) is a standard assumption
for most astrophysical models.
These two criteria imply that
\begin{equation}\label{nnrho}
0\leq p\leq \rho\ \mbox{and}\ |j| \leq \rho,
\end{equation}
cf.\ \cite{HE} and \cite{P}.
Furthermore, by (DEC) any world line $(s,R(s))$ of a material particle
or a photon satisfies the estimate
\[
\left|\frac{dR}{ds}(s)\right| \leq e^{(\mu-\lambda)(s,R(s))}
\]
so that locally the speed of energy flow is
less than or equal to the speed of light. In addition we need
to assume certain a-priori information on the behavior of the
solutions of the Einstein-matter equations, namely that
for solutions launched by data from a suitable class
\begin{itemize}
\item
$\gamma^+$ defined by (\ref{gamma+}) exists on $[0,\infty[$, and
the solution exists on $D$ \hspace*{\fill}(GLO)
\item There exists a constant $c_1>0$ such that
$\rho \leq -c_1 j$ in $D$. \hfill(GCC)
\end{itemize}
The role of the ``global existence condition'' (GLO) is obvious.
The ``gravitational collapse condition'' (GCC) is crucial for our 
method of proof, in particular, together with (\ref{nnrho})
it implies that $j\le 0$ in $D$, i.e., the matter is ingoing for all times.
Notice also that in (\ref{dthatmu}) condition (GCC) was used.
We emphasize that for Vlasov matter the conditions (DEC) and (NNP)
hold always, while (GLO) and (GCC) follow via (\ref{west})
from a suitable restriction of the initial data.

For a general matter model satisfying these assumptions
not all the results in Theorem~\ref{bh} can be obtained
but the following still holds:
\[
\lim_{s\to\infty}\gamma^+(s) < \infty
\ \mbox{and}\ \lim_{t\to \infty} \mu(t,r) = -\infty\ \mbox{for}\
\lim_{s\to\infty}\gamma^+(s) \leq r \leq r_1
\]
for some $r_1 > \lim_{s\to\infty}\gamma^+(s)$. If
$r^\ast$ and $\gamma^\ast$ are defined as in Theorem~\ref{bh}
then
\[
\lim_{s\to\infty}\gamma^\ast (s) < \infty,
\]
and every  radially outgoing null geodesic $\gamma$ with $\gamma(0) > r^\ast$
is future complete with $\lim_{s\to\infty}\gamma (s) = \infty$.
The analysis in Section~\ref{subsecbh} applies to the resulting spacetimes
as well, in particular since they again are vacuum for $r\geq R_1$.
\subsection{Related results}
Due to the inherent difficulties of the Einstein field equations
progress towards understanding the issues of cosmic censorship
and the formation of black holes has up to now been restricted
to the case of spherical symmetry. At least this assumption is made
in all the papers we mention below.

The most complete understanding of the issues at hand has so far 
been obtained for the case where the Einstein equations are coupled 
to a massless scalar field, cf.\ 
\cite{Chr86,Chr87,Chr91,Chr93,Chr94,Chr99a,Chr99}.
The final result of these investigations is that weak and strong
cosmic censorship hold for the Einstein-scalar field system.
A crucial step was the investigation in \cite{Chr91} where explicit
conditions on the initial data were formulated which guarantee the
formation of a trapped surface. Although the conditions we impose
for Theorem~\ref{bh} are reminiscent of the ones in \cite{Chr91},
there is also an important difference. In \cite{Chr91} the admissible
data cover the full range of $2 m / r \in ]0,1[$, whereas for our result
we need that initially $2 m / r$ is close to one in the outer matter
ring. 

However, there may be a better reason for this difference than just
the limited abilities of the present authors. The spherically symmetric
Einstein-Vlasov system can exhibit at least the following qualitatively 
different types of behavior. Firstly, for small data the solution disperses
in the sense of Theorem~\ref{glex}. Secondly, the system has a tremendously 
rich family of steady states, cf.\ \cite{Rss1,RRss1,RRss2} and also
\cite{AR2} concerning the possible shapes these steady states can take.
Numerical investigations show clearly that there are both stable and
unstable steady states, cf.\ \cite{AR1}. 
There is also strong numerical evidence \cite{AR1} 
that the system has time-periodic
or at least almost periodic solutions, which is the third type of solution
behavior. The fourth and last type of solution behavior is gravitational
collapse and formation of a black hole as shown by Theorem~\ref{bh}.
To our knowledge only dispersion and gravitational collapse are
known as possible solution behaviors for the Einstein-scalar field system,
and this wider range of qualitative solution behaviors for the Vlasov matter
model may explain why the condition needed to force gravitational collapse
is more restrictive than for the scalar field.

In passing we also note that the primary motivation for coupling the
Einstein equations to a scalar field is, according to \cite{Chr99},
to capture the wave nature of non-symmetric vacuum solutions to the Einstein
equations in a spherically symmetric set up while still enjoying
the simplifications which the latter symmetry assumption entails.
As mentioned above, the Vlasov matter model is actually used in
astrophysics to describe galaxies or globular clusters.

Another matter model which features prominently in the history
of the concepts of gravitational collapse and black hole is
a fluid with pressure zero which is usually termed dust.   
The analysis in \cite{OS} has definitely shaped the
overall picture of gravitational collapse to a large extent.
The physical reason for neglecting pressure is the
intuition that once the collapse is sufficiently advanced,
gravity will dominate all other forces, including pressure.
One important mathematical advantage of this matter model
is that one can use coordinates which are co-moving with the
matter. On the other hand, dust produces shell-crossing
singularities which hinder the analysis of the real
issues, and dust can lead to naked singularities as
shown in \cite{Chr84}. Whether cosmic censorship will eventually
be established for Vlasov matter or not, solutions of the
Einstein-Vlasov system do have non-trivial pressure,
and the system has so far not been shown to produce naked
singularities.

In making the last statement we are fully
aware of \cite{ST}. There it is claimed that numerical simulations
of the axially symmetric Einstein-Vlasov system can lead to naked
singularities. However, there is reason to believe that the
matter model which was actually simulated in \cite{ST} was dust and
not Vlasov, cf.\ \cite{Rend92}.

To conclude this comparison with previously known results we mention
that in \cite{Rend92} a continuity argument was used to show that there
exist initial data for the spherically symmetric Einstein-Vlasov system
which lead to the formation of trapped surfaces. Combining this with
the results of \cite{D05} and \cite{DR05} implies that there
exist data which lead to the formation of a black hole. Due to the method
of proof in \cite{Rend92} this analysis does not give explicit conditions
on the data which guarantee this type of behavior. Our approach
does produce such explicit conditions on the data, and these data
are stable against perturbations which are small in a suitable sense. 
 
\subsection{Open problems and future perspectives}\label{secpers}

To conclude we mention some open problems and possible perspectives
for future research. An immediate question is whether the geodesic
$\gamma^\ast$ in Theorem~\ref{bh} is future complete;
the question whether an event horizon has complete generators
is of general interest.
This issue is closely related to the question whether in the
limit $t\to \infty$ all the matter ends up in the region $\{r \leq 2 M\}$,
i.e., whether $\lim_{t\to\infty} m(t,2M) = 2 M$. It is clear from
the geodesic equation for $\gamma^\ast$ that this geodesic can not be complete
if it is running in vacuum all the time, cf.\ (\ref{rddotinvac}), 
which means that at least some matter
must cross $r=2 M$ if $\gamma^\ast$ should be complete.
These questions are currently under investigations by two of the authors,
and we believe that for suitably restricted data indeed
$\lim_{t\to\infty} m(t,2M) = 2 M$. 

The next logical step will be to analyze the situation corresponding
to Theorem~\ref{bh} in a coordinate system which has the
potential to cover regions of spacetime containing trapped surfaces
and which may reach all the
way to the spacetime singularity which forms at the center. 
Preliminary steps in this direction look promising.

Much more demanding is the question of how to relax the conditions
in Theorem~\ref{bh} so that the numerical
observations reported in \cite{AR1} where the perturbation of 
unstable steady states leads to the formation of black holes
are covered. An analytic understanding of the stability properties
which were observed numerically is a further open problem.
An answer to this could also help to explain the phenomenon
of critical collapse, cf.\ \cite{AR1,OC,RRS2} for relevant
numerical results for the Einstein-Vlasov system and \cite{Gund}
for a general discussion of this issue.

As pointed out above the question whether weak cosmic censorship
holds for the spherically symmetric Einstein-Vlasov system is
related to the question whether this system has global
solutions in Schwarzschild coordinates or not.
Useful estimates in the direction of global existence in Schwarzschild
coordinates, which go beyond what was 
reported in Section~\ref{seclocglob}, were established in
\cite{And1}, but the problem remains open. And when thinking about 
weak cosmic censorship one should definitely keep the possibility
in mind that global existence in Schwarzschild coordinates
could be violated for an initial data set ``of measure zero'',
while weak cosmic censorship could still be true for
the spherically symmetric Einstein-Vlasov system.

Eventually one will wish to go beyond spherical symmetry,
and any extension of the results mentioned in these notes
to for example the case of axial symmetry is in our opinion
a challenging and worthwhile problem.

\smallskip

\noindent
{\bf Acknowledgement}.
The authors would like to thank Piotr Chru\'{s}ciel, Helmut Friedrich,
and Alan Rendall for valuable discussions.

\end{document}